\documentclass[12pt]{iopart}

\usepackage{graphicx}
\usepackage{iopams}

\begin{document}

\title{Low-density, one dimensional quantum gases in the presence of a
  localised attractive potential}

\author{J.~Goold, D.~O'Donoghue, and Th.~Busch}
\ead{jgoold@phys.ucc.ie}

\address{Department of Physics, National University of Ireland, UCC,
  Cork, Ireland}

\begin{abstract}
  We investigate low-density, quantum-degenerate gases in the presence
  of a localised attractive potential in the centre of a
  one-dimensional harmonic trap. The attractive potential is modelled
  using a parameterised $\delta$-function, allowing us to determine
  all single particle eigenfunctions analytically. From these we
  calculate the ground state many-body properties for a system of
  spin-polarised fermions and, using the Bose-Fermi mapping theorem,
  extend the results to strongly interacting bosonic systems. We
  discuss the single particle densities, the pair correlation
  functions, the reduced single particle density matrices and the
  momentum distributions as a function of particle number and strength
  of the attractive point potential. As an important experimental
  observable, we place special emphasis on spatial coherence
  properties of such samples.
\end{abstract}


\maketitle

\section{Introduction}
\label{sec:Introduction}

The rapid advances made in the fields of atom cooling and trapping
have created re-newed theoretical and experimental interest into
lower-dimensional systems \cite{Bloch:07,Prentiss:99,Ketterle:01,
  Zimmermann:01,Birkl:02,Raithel:02,Esslinger:05}. Even though
restricting the spatial degrees of freedom often leads to stronger
correlation, various exactly solvable models are known that cover
different temperature and interaction regimes. One example of this is
the Lieb-Liniger gas \cite{Liniger:63,Lieb:63}, a model for a quantum
gas of bosons trapped in one dimension and interacting via a
point-like potential. Using the Bethe ansatz \cite{Bethe:31}, this
model can be exactly solved and it was realised by Girardeau
\cite{Girardeau:60} that in the limit of infinitely strong repulsive
interactions a particularly elegant solution can be found by mapping
it onto a gas of free fermions.

This strongly correlated limit of the Lieb-Liniger model is known as
the Tonks-Girardeau (TG) gas \cite{Girardeau:60,Tonks:36} and the
first examples for which it was solved were free space and box-like
systems with periodic boundary conditions
\cite{Girardeau:60,Girardeau:65}. Recently quantum gases in the TG
limit were experimentally realised \cite{Paredes:04,Weiss:04} and used
to observe non-equilibrium dynamics \cite{Kinoshita:06}. In
anticipation and in light of the experimental realisation Wright and
Girardeau managed to extend the number of exactly known solutions by
describing the gas subject to a harmonic trap
\cite{Wright:00,Triscari:01} and several other authors have considered
a range of different potentials
\cite{Huyet:03,Fu:06,Wu:07,Goold:08,Yin:08}. In this work we extend
the recent progress and explore the ground state properties of a TG
gas (and by means of the mapping theorem also of a gas of
non-interacting, spin-polarised fermions) in a harmonic trap in the
presence of a localised attractive potential at the trap centre. We
place particular emphasis on spatial coherence effects.

Even though a description for solving the TG gas analytically exists
it is often still necessary to involve numerical calculations,
especially if a larger number of particles is considered. While these
can be time and resource consuming, approximations have been found
recently to investigate ground state properties of samples with large
particle numbers \cite{Forrester:03,Papenbrock:03}. For mesoscopic
samples an algorithm allowing efficient calculation of the reduced
single particle density matrix (RSPDM) for a TG gas in an arbitrary
potential was recently presented by Pezer {\sl et al.}
\cite{Pezer:07}. As many of the important groundstate properties of a
many-body systems can be directly calculated from a system's reduced
single particle density matrix we will employ this algorithm in this
work.

Localised, attractive potentials have recently been used with and
suggested for several experiments within ultracold quantum
gases. Highly focussed optical beams were shown to allow an increase
in the phase-space density of a gas and drive its transition towards
Bose-Einstein condensation \cite{StamperKurn:98,Weber:03}. Uncu {\sl
  et al.} have shown that such processes can be analysed using
point-like functions \cite{Uncu:07,Uncu:08}. A second way of creating
a highly localised potential is given by trapping an impurity inside a
cloud of cold atoms. For ions, for example, the trapping frequencies
can be several $100$ kHz and they therefore provide very localised
potentials \cite{Cote:02}. The harmonic trap with a point like
potential at the center as considered in this work is a well fitting
toy model for such a situation.

In order to find the many-body solutions in a given geometry for the
TG gas (or for non-interacting fermions) one must know the exact
single particle eigenstates. This is a problem in itself, as the list
of exactly solvable single particle problems in quantum mechanics is
limited. The system we consider here is the harmonic trap with a
point-like potential trapped at its centre. For a repulsive potential
this resembles the limit of a double well trap and was recently
investigated for boson as well as fermionic systems
\cite{Huyet:03,Goold:08}. The groundstate physics for the same
potential of a bosonic pair, up to and including the Tonks regime, was
also rigorously analyzed \cite{Murph:07}. Here we describe the systems
properties for the other limit, i.e.~for a trap with an attractive
central $\delta$-potential. It is important to emphasize that this is a
setting very different to the repulsive case as the attractive
potential possesses an additional bound state.

Through the Fermi-Bose mapping theorem the bosonic wave-function can
be calculated directly from the appropriately chosen fermionic one by
symmetrization. Since the symmetry or antisymmetry of a wave function
does not have an influence on the density distribution, the spatial
density profiles for fermionic and bosonic samples in this limit are
indistinguishable. Therefore, whenever results concerning density
distributions are presented in this paper, they apply to fermionic as
well as to bosonic samples.

The paper is organised as follows: in Sec.~\ref{sec:ModelHamiltonian}
we define the many-body Hamiltonian for our system and briefly describe
the technique to solve it for non-interacting fermions as well as for
hard-core bosons. In Sec.~\ref{sec:EigenstatesDeltaSplitTrap} we
review the single particle eigenspectrum of the the harmonic trap with
a central attractive $\delta$-function and in
Sec.~\ref{sec:FermiBoseGSP} we calculate the many-body groundstate
properties of a TG gas in such a potential. The corresponding results
for free fermions are also shown. Special emphasis is put on spatial
coherence effects. Finally we conclude in Sec.~\ref{sec:Conclusions}.

\section{Model Hamiltonian}
\label{sec:ModelHamiltonian}

We consider a gas of $N$ identical atoms trapped in a tight atomic
waveguide, such that the dynamic of the gas is strongly restricted in
the transversal directions. In the low-temperature limit this allows
us to choose a one-dimensional model where the parameterisation of the
scattering interaction takes the three dimensional nature of the
particles into account \cite{Olshanii:98}. In the remaining direction
we then consider a harmonic potential perturbed in the center by a
well localised attractive potential, which we model by a parameterised
$\delta$-function. For sufficiently low density we only need to
consider two-particle collisions and the Hamiltonian can the be
written as
\begin{equation}
  \label{eq:Hamiltonian}
  \mathcal{H}=\sum_{n=1}^N
              \left(-\frac{\hbar^2}{2m}\frac{\partial^2}{\partial x^2}
                    +\frac{1}{2}m\omega^2x_n^2-\kappa\delta(x_n)\right) 
              +\sum_{i<j}V(|x_i-x_j|)\;,
\end{equation}
where $m$ is the mass of a single atoms, $\omega$ the frequency of the
harmonic potential and $\kappa$ the strength of the point-like
attractive potential, which is located at $x_n=0$.  The
particle-particle interaction potential depends only on the relative
distance between two particles.

For purposes of clarity and readability we re-scale the above
Hamiltonian to harmonic oscillator units where all length are in units
of the ground state size, $a_0=\sqrt{\hbar/m\omega}$, and all energies
in terms of the oscillator energy, $\hbar\omega$,
\begin{equation}
 \label{eq:ham_scaled}
   \bar{\mathcal{H}}=\sum_{n=1}^N
            \left(-\frac{1}{2}\frac{\partial^2}{\partial \bar{x}^2}
                  +\frac{1}{2} \bar{x}_n^2-\bar{\kappa}\delta(\bar{x}_n)\right) 
              +\sum_{i<j}V(|\bar{x}_i-\bar{x}_j|)\;.
\end{equation}
This leads to a new scaled strength for the attractive potential given
by $\bar{\kappa} = ( \hbar \omega a_0)^{-1} \kappa$. For notational
simplicity we shall drop the overbars on all scaled quantities and
acknowledge that we are, henceforth, dealing in the scaled units just
described. All units used in calculations and figure plots in this
paper are in terms of these scaled units.

\subsection{Fermions}

In order to avoid any confusion we first consider the interaction term
in the bosonic and fermionic cases separately. Fermionic systems can
be analyzed by realizing that due to the antisymmetric nature of the
wave-function no $s$-wave scattering between two particles can
happen. As $p$-wave scattering at low temperature is negligible (unless
close to a resonance), the Hamiltonian (\ref{eq:Hamiltonian}) can be
approximated by neglecting any inter-particle scattering and solving
the problem for free particles.

\subsection{Bosons}
For bosonic systems at low temperatures the atomic interaction
potential can be well approximated by a point-like potential,
$V(|x_i-x_j|)=g_{1D}\delta(|x_i-x_j|)$. This approximation is
frequently used and the only reminiscence of the exact potential is
given by the three-dimensional $s$-wave scattering length, $a_{3D}$
\cite{Huang:57}. For positive values of $a_{3D}$ the interaction is
repulsive and for negative values it is attractive. This scattering
length is then related to the one-dimensional coupling constant via
\begin{equation}
  g_{1D}=\frac{4\hbar^2 a_{3D}}{ma_\perp}
         \left(a_\perp-Ca_{3D}\right)^{-1}\;,
\end{equation}
where $C$ is a constant of value $C=1.4603\dots$
\cite{Olshanii:98}. In this work we will purely focus on very strong
repulsively interacting systems, which in low dimensions corresponds
to the low density limit and is known as the Tonks-Girardeau limit.

One of the remarkable features of this limit is that it becomes
exactly solvable using the so-called Fermi-Bose mapping theorem. The
theorem follows from replacing the interaction term in the Hamiltonian
in eq.~(\ref{eq:ham_scaled}) with the following constraint on the
allowed wave-functions
\begin{equation}
  \label{eq:constraint}
  \Psi=0\qquad if \quad |x_i-x_j|=0,\qquad i\neq j\quad(1\leq i\leq\ j\leq N)\;,
\end{equation}
which is equivalent to the demands of the Pauli-exclusion principle
for a gas of spinless fermions. One can therefore compute the
many-body bosonic ground-state wave-function from the fermionic case,
using the simple symmetrization
\begin{equation}
  \label{eq:wavefunc}
  \Psi_B(x_1,.....x_N)=|\Psi_F(x_1,.....x_N)|\;.
\end{equation}
This procedure transforms the strongly interacting bosonic problem
into a problem treating non interacting fermions. As for
non-interacting fermions calculation tools are known, the problem
becomes treatable.

Therefore the bosonic as well as the fermionic problem can be solved
if the single particle problem can be solved. If this solution is even
analytic, no further approximations to the many-particle wave-function
have to be made.

\section{Eigenstates of the harmonic trap with central attractive
  point potential}
\label{sec:EigenstatesDeltaSplitTrap}

\begin{figure}
\centering
\includegraphics[width=.5\linewidth]{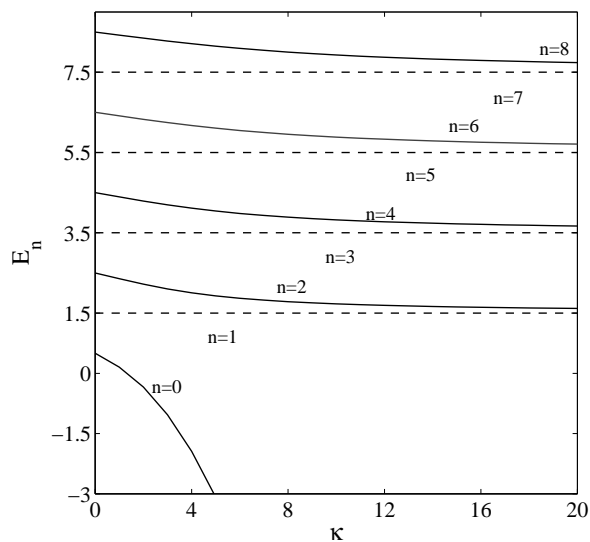}
\caption{Single particle energy eigenspectrum for the harmonic trap
  with an attractive point potential at the trap center. The solid
  curves correspond to the symmetric eigenstates and the broken lines
  to the anti-symmetric eigenstates. Note the existence of the trapped
  eigenstate originating from the attractive point-potential.}
  \label{fig:EnergyEigenSpectrum}
\end{figure}
Even though the eigenstates of a harmonic trap with a central
attractive $\delta$-potential are well known \cite{Busch:98}, we will
briefly review them in this section. It is easy to see that owing to
its point-like nature all eigenenergies associated with odd eigenstates
of an undisturbed harmonic oscillator are independent of the strength
of the $\delta$-function.  All energies associated with the even
eigenstates, on the other hand, will decrease non-trivially with
increasing values of $\kappa$. And while all excited even eigenstates
are bound from below by the energies of the next, lower lying odd
eigenstate, the energy of the ground state $E_0$ becomes unbounded
from below in the $\kappa\rightarrow\infty$ limit (see
Fig.~\ref{fig:EnergyEigenSpectrum}). This is due to the existence of a
bound state within the attractive point potential which is accessed by
the system once the ground state energy $E_0<0$, which corresponds to
$\kappa = 0.675978$. The even eigenstates are then given by
\begin{equation}
 \label{eq:sin_part_sym_vec}
 \psi_n(x)=\mathcal{N}_n\; e^{-\frac{x^2}{2}} 
           U\left(\frac{1}{4}-\frac{E_n}{2},\frac{1}{2},x^2 \right) 
           \qquad n=0,2,4\ldots\;,
\end{equation}
where $\mathcal{N}_n$ is the normalization constant and $U(a,b,z)$ are
the Kummer functions \cite{abr72}. The corresponding eigenenergies,
$E_n$, are determined by the roots of the implicit relation,
\begin{equation}
 \label{eq:sin_part_sym_val}
\kappa = 2 \frac{\Gamma\left( -\frac{E_n}{2} + \frac{3}{4}\right)}
                  {\Gamma\left( -\frac{E_n}{2} + \frac{1}{4}\right)}\;.
\end{equation}
By contrast, the antisymmetric
eigenfunctions vanish at the origin and are unaffected by the point potential.
They are therefore given by the odd eigenstates of the unperturbed
harmonic potential ($\kappa =0$)
\begin{equation}
  \label{eq:sin_part_anti_vec}
 \psi_n(x) = \mathcal{N}_n H_n (x) e^{-\frac{x^2}{2}} \quad n =1,3,5\ldots\;,
\end{equation}
where $H_n(x)$ is the $n^{th}$ order Hermite polynomial.  The
corresponding energies are given by the eigenvalues of the odd parity
states of the harmonic oscillator, $E_n = \left( n + \frac{1}{2}
\right)$. Since we know the single particle eigenstates, we can build
the Slater determinant for the problem of non-interacting fermions and
through the symmetrisation of the mapping theorem we can calculate the
exact many-body bosonic wave-function.

\section{Ground-state properties}
\label{sec:FermiBoseGSP}

\subsection{Single-particle densities and pair-distribution functions}
\label{subsect:spd}
\begin{figure}[tbp]
  \includegraphics[width=\linewidth]{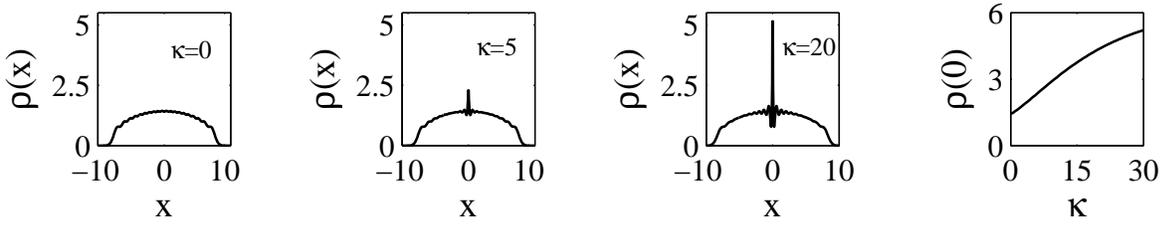}
  \caption{The first three plots from left to right show the single
    particle densities $\rho(x)$ for 20 particles and for different
    values of the potential strengths $\kappa=0, 5$ and $20$. As the
    attractive potential is assumed to be point-like, it has no
    significant influence on the size of the atomic cloud and only
    introduces a very localised disturbance. The height of the central
    disturbance, $\rho(0)$, is shown as a function of $\kappa$ in the
    right-most plot.}
 \label{fig:DensityN20}
\end{figure}

The single particle density for the bosonic as well as the fermionic
system is defined as
\begin{equation}
  \label{eq:density}
  \rho(x)=N \int_{-\infty}^{+\infty}|\Psi_B(x,x_2\dots,x_N)|^2dx_2\dots dx_N
         =\sum_{n=0}^{N-1} |\psi_n(x)|^2\;.
\end{equation}
In Fig.~\ref{fig:DensityN20} we show $\rho(x)$ for a gas of 20
particles for three different values of $\kappa$. The introduction of
the attractive point potential to the harmonic trap results in the
emergence of a central density spike, which grows quasi-linearly with
increasing strength of the attractive potential (see
Fig.~\ref{fig:DensityN20}, right most figure). This feature originates
from the strongly localised nature of the bound state of the
$\delta$-function and, which is also the reason for the potential
having virtually no influence on the overall width of the density
distribution.

\begin{figure}[tb]
  \includegraphics[width=\linewidth]{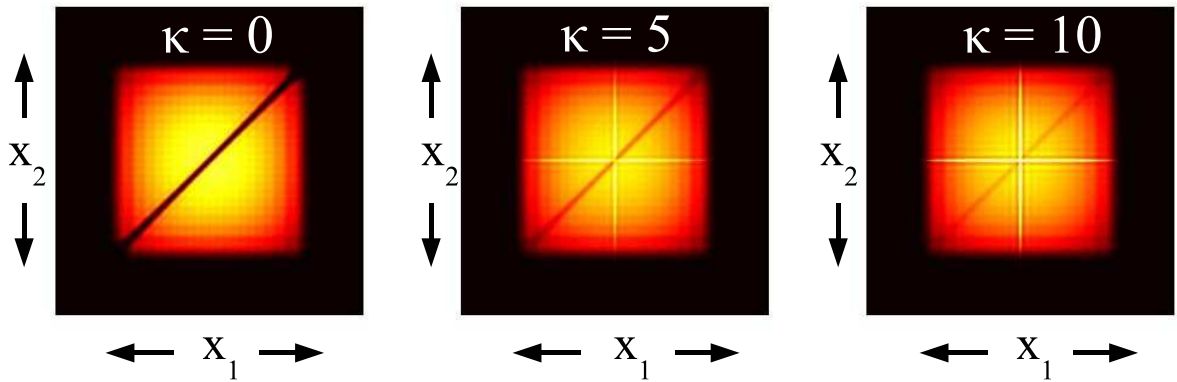}
  \caption{Pair-distribution function $D(x_1,x_2)$ for $N=20$
    particles and potential strengths $\kappa=0, 5$, and $10$.  Each
    plot spans the range $ -10 < x_1, x_2 < 10$. The manifestation of
    the bound state within the attractive potential can be clearly
    seen in the emergence of a cross of localised probability along
    the lines $x_1,x_2=0$ }
 \label{fig:4by4pdf}
\end{figure}

The pair-distribution function, $D(x_1,x_2)$, is a two particle
correlation function that describes the probability to measure two
different particles at two given positions at the same time. It is
defined in the following way
\begin{eqnarray}
  \label{eq:pdf}
    D(x_1,x_2)&=&N(N-1)\int_{-\infty}^{+\infty} 
                 |\Psi_B(x_1,x_2\dots,x_N)|^{2} dx_3 \dots dx_N\;,\\
              &=&\sum_{0\leq n\leq n'\leq N-1}^{N-1} 
                 |\psi_n(x_1) \psi_{n'}(x_2)-\psi_n(x_2) \psi_{n'}(x_1)|^2\;,
\end{eqnarray}
and in Fig.~\ref{fig:4by4pdf} we show its evolution for a sample of 20
particles under increasing strength of the point potential. Most
notably, in the $\kappa=0$ case we notice the absence of any
probability along the line $x_1=x_2$, which is due to the strong
repulsion between the particles.  Secondly, with increasing $\kappa$ a
cross type pattern of high probability density emerges along the
$x_1=0$ and $x_2=0$ lines.  This is in agreement with the appearance
of the peak at $x=0$ in the density plots and corresponds to the
lowest eigenstate becoming bound and strongly localized within the
point potential. Again, the appearance of the localized disturbance
does not have any large influence on the pair distribution function at
larger values of $x$.

\subsection{ Reduced single particle density matrices and natural
  orbitals}
\label{subsect:rspd} 

Although a many-body wave-function fully characterizes a quantum
mechanical state, a RSPDM is a useful and convenient tool for deriving
many important properties of many-body systems. In particular the
expectation values of many important one-body quantities such as the
momentum distribution or the von Neumann entropy are easily obtained
from it. The RSPDM for a 1-D gas of spinless, non-interacting fermions
is given by
\begin{equation}
  \label{eq:rdspdmf}
  \rho_F(x,x')=\sum_{n=1}^N\psi_n(x) \psi^*_n(x')\;,
\end{equation}
and is diagonal by default since it is a projector onto an
$N$-dimensional subspace of the Hilbert space of possible one particle
states. Although we do not show plots of the matrices here, we will
refer to the RSPDM later when we investigate how the momentum
distribution of a harmonically trapped Fermi sea is modified by
introduction of the attractive point potential.

The RSPDM for the bosonic case is defined as
\begin{equation}
  \label{eq:rspdm_def}
  \rho_B(x,x')=\int_{-\infty}^{+\infty} 
            \Psi^B_0(x,x_2,\dots,x_N)\Psi^B_0(x',x_2,\dots,x_N)dx_2\dots dx_N\;,
\end{equation}
and normalized to $\int \rho_B(x,x) dx=N$. Due to the inter-bosonic
interactions it is not diagonal in the basis of single particle states
and for finite values of $\kappa$ we will solve for it numerically
below. A naive calculation strategy is still a numerical feat due to
the large demands on memory space. However, Pezer and Buljan
\cite{Pezer:07} have recently presented an algorithm that allows this
calculation to be carried out very effectively and it is this
algorithm which we employ here to calculate $\rho_B(x,x')$. Note that
for the $\kappa=0$ case the integral (\ref{eq:rspdm_def}) was recently
solved analytically \cite{Lapeyre:02}.

\begin{figure}[tbp]
  \includegraphics[width=\linewidth]{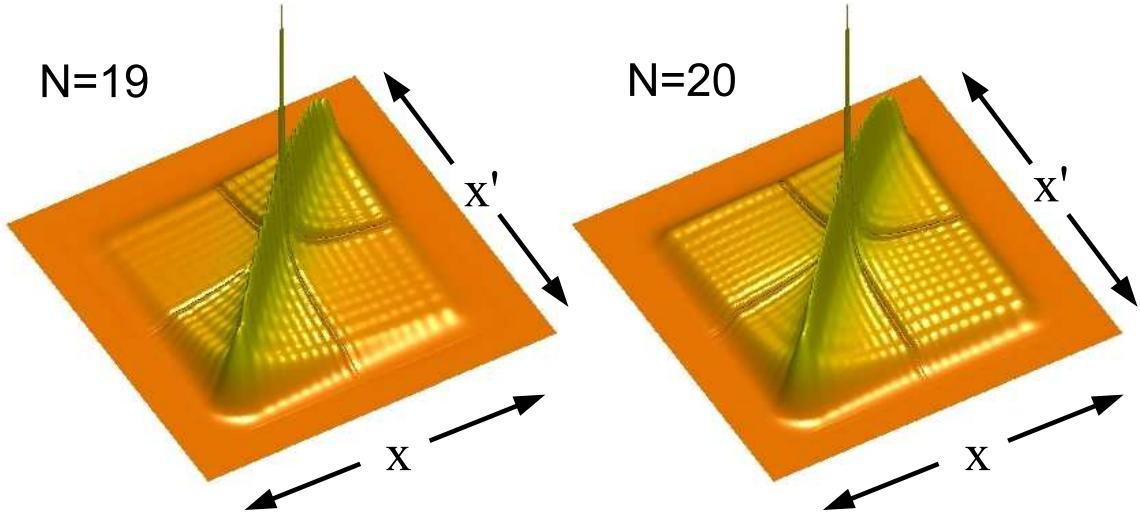}
  \caption{RSPDM $\rho_B(x,x')$ for $N=19$ and $N=20$ bosons in a
    harmonic trap with a central attractive point potential of
    strength $\kappa=10$. One notices a distinct difference in the
    off diagonal behavior for different particle numbers. Each
    plot spans the range $ -10 < x, x' < 10$. }
 \label{fig:rspdm4byf}
\end{figure}

The RSPDM expresses self correlation and one can view $\rho_B (x,x')$
as the probability that, having detected the particle at position $x$,
a second measurement, immediately following the first, will find the
particle at the point $x'$. Classically one would only expect a result
for $x=x'$, however quantum mechanically off-diagonal correlations
become important. In Fig.~\ref{fig:rspdm4byf} we show the RSPDM for an
odd ($N=19$) and an even ($N=20$) number of particles in the presence
of a strong attractive potential with $\kappa=10$. In both figures a
spike at $x=x'=0$ is the dominating feature, originating from the
bound state within the delta-function and matching with the results
for the single particle density (\ref{eq:density}), which can be
obtained from the diagonal $\rho(x)=\rho_B(x,x'=x)$ . This localised
nature also explains the absence of any probability density in the
cross defined by the lines $x = 0, x' = 0$.  It is clear from
Fig.~\ref{fig:rspdm4byf} that there is a distinct difference between
systems with odd and even particle numbers. In the $N=19$ case we see
that the probability in the off-diagonal quadrants is strongly
depleted as compared to the $N=20$ system. This effect occurs for all
consecutive odd and even particle numbers studied. A similar odd-even
effect was observed in the repulsively split trap \cite{Goold:08} and
later in the split box \cite{Yin:08}. Contributions in the off
diagonals of the RSPDM mean that the sample has some degree of spatial
coherence and we will interpret this effect with respect to physically
observable quantities in the next section.

\begin{figure}[tbp]
  \includegraphics[width=\linewidth]{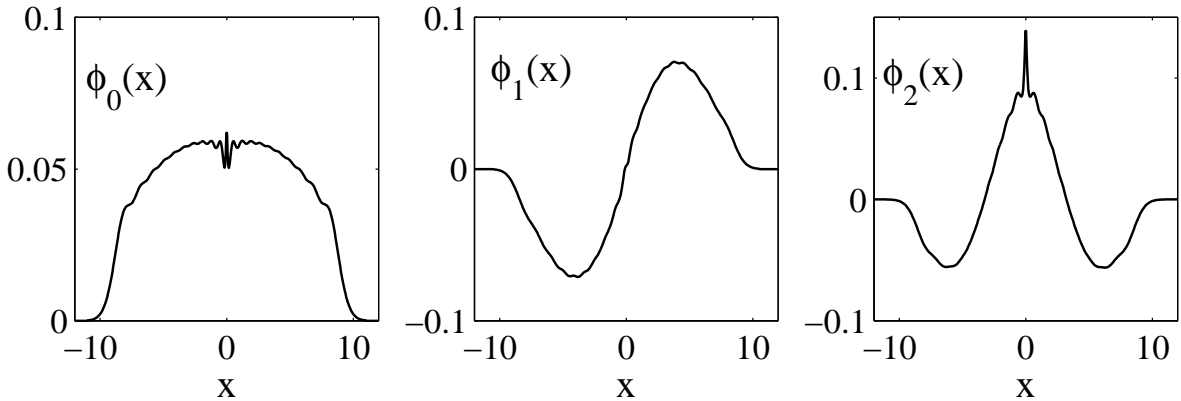}
  \caption{The first three, energetically lowest lying natural
    orbitals for a Tonks-Girardeau gas of 20 particles in a harmonic
    trap with an attractive point potential of strength
    $\kappa=10$. The influence of the attractive potential on the even
    states can be clearly seen.}
 \label{fig:naturalo}
\end{figure}

In order to allow for an easier interpretation of the bosonic RSPDM we
change to a representation in which the matrix becomes diagonal
\begin{equation}
  \label{eq:norbs}
  \int_{-\infty}^{\infty}\rho_B(x,x') \phi_j(x') dx'=\lambda_j \phi_j(x)\;.
\end{equation}
The eigenfunctions $\phi_j(x)$ are known in theoretical chemistry as
'natural orbitals' and their associated eigenvalues $\lambda_j$
represent the occupation number of the eigenvector. The first three
natural orbitals with lowest energy of a 20 particle harmonically
trapped TG gas in the presence of an attractive point potential,
$\kappa=10$, are shown in Fig.~\ref{fig:naturalo}. One can see that
the point potential has a strong, localised influence on the shape of
the symmetric orbitals, whereas the antisymmetric orbitals are not
affected and retain the same shape as in the $\kappa=0$ case
\cite{Triscari:01}.

\subsection{Ground State Occupation Numbers}
\label{subsect:occ}

The largest macroscopic eigenvalue $\lambda_0$ measures the fraction
of particles that are in the $\phi_0(x)$ orbital, sometimes known as
the 'BEC' state, by $f=\frac{\lambda_0}{N}$. It can hence be used as a
measure of the coherence in the system. For the TG gas this was first
studied by Girardeau {\sl et al.} for the simple harmonic trap
~\cite{Triscari:01} and small particle numbers. They found that
despite the strong interactions macroscopic coherence effects can
still exist.  Later Forrester {\sl et al.}  showed that as one
increases the particle number, $\lambda_0$ tends toward $\sqrt{N}$
\cite{Forrester:03}. Here we study how increasing the strength of the
attractive point-like potential affects this $\sqrt{N}$ behavior.

\begin{figure}[tbp]
\includegraphics[width=\linewidth]{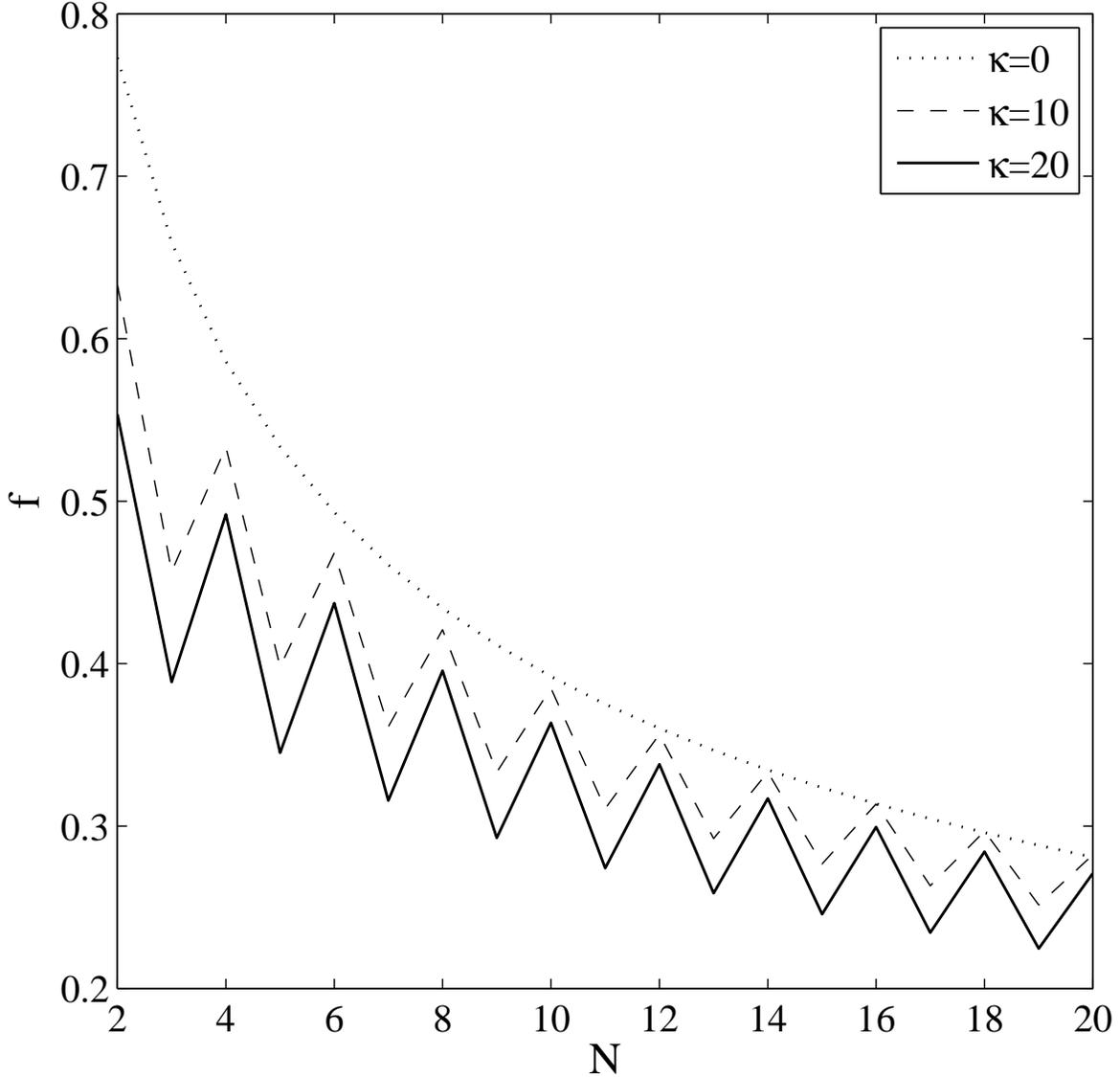}
\caption{ The ground state occupation fraction $f=\frac{\lambda_0}{N}$
  as a function of particle number $N$ for the potential strengths $
  \kappa=0,10$ and $20$.}
\label{fig:occ}
\end{figure}

The groundstate occupation fraction, as a function of $N$, for the
bosonic TG gas is shown in Fig.~\ref{fig:occ} for two different
potential strengths and compared to the simple harmonic oscillator
result. One can see that the introduction of the central potential
creates a distinct, oscillating pattern, which become more pronounced
for increased depth. We notice that when the particle number is odd
the value of $\lambda_0$ or the 'coherence' is relatively lower than
when the particle number is even. These oscillations in the occupation
fraction damp out with increasing number of particles and are similar
to an effect recently found for a repulsive $\delta$-barrier
\cite{Goold:08} (see also \cite{Yin:08}). The main difference is a
reversal of the oscillations, with the even number samples coherence
being decreased stronger in the repulsive case.

This oscillation pattern was identified as being the result of a
pairing of single particle energy levels in the repulsive case, which
is also the explanation in this situation. While, however, in the case
for $\kappa>0$ consequent even energy levels increased in energy and
paired with the next higher lying odd levels, here consequent odd-even
levels pair due to a decrease in the energy of the even levels (see
Fig.~\ref{fig:EnergyEigenSpectrum}). At the same time the remaining
ground state is becoming unbounded from below. A more qualitative
picture can be given by assuming one-particle to be strongly bound
within the $\delta$-potential and acts as a barrier for the remaining
$N-1$ particle system.

Let us in the following investigate how this odd-even coherence effect
manifests itself in two experimentally realizable quantities, namely
the momentum distribution and interference fringe visibility during
free temporal evolution.

\subsection{Momentum Distribution}
\label{subsect:mom}

\begin{figure}[tbp]
  \includegraphics[width=\linewidth]{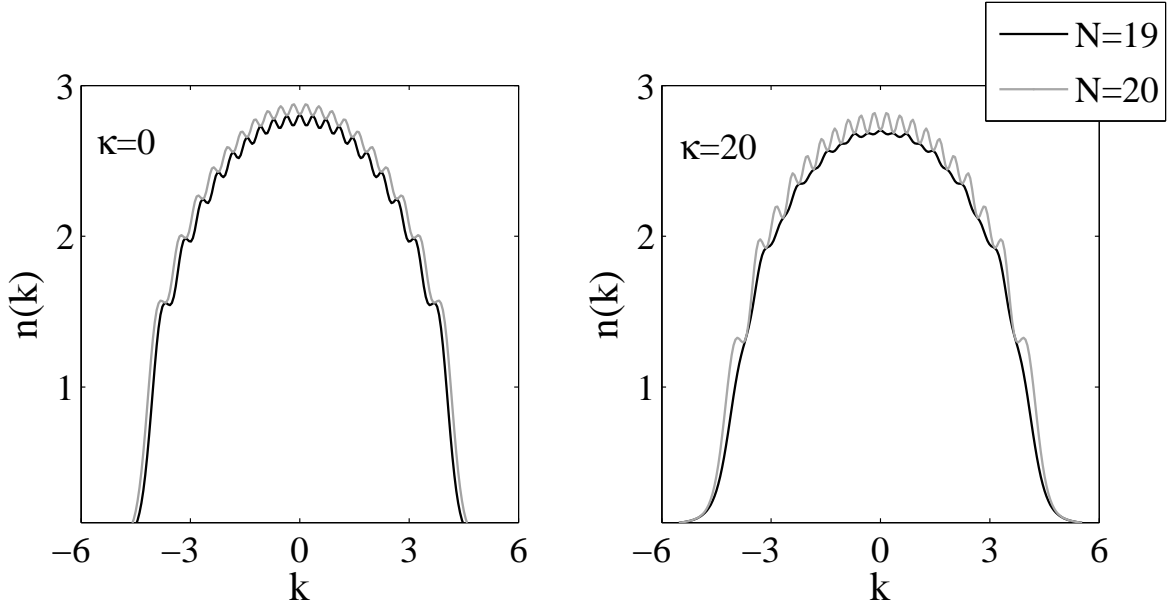}
  \caption{Momentum distributions of a $N=19$ (black) and a $N=20$
    (grey) particle spin-polarized Fermi gas in a harmonic trap with
    central attractive point potential. The $\kappa=0$ and $\kappa=20$
    cases are shown separately.}
 \label{fig:momf}
\end{figure}

Although the spatial density profiles are equivalent for a gas of
non-interacting fermions and strongly interacting bosons they still
show distinctly different momentum distributions
\begin{equation}
  \label{eq:momentum}
  n_{B(F)}(k)=(2\pi)^{-1}\int_{-\infty}^{\infty}dx\int_{-\infty}^{\infty}dx'
  \rho_{B(F)}(x,x') e^{-ik(x-x')}\;,
\end{equation}
where the normalization is chose to be
$\int_{-\infty}^{\infty}n(k)dk=N$. Alternatively, the spectral
decomposition of $\rho_{B(F)}(x,x')$ allows us to compute the momentum
distribution for arbitrary particle number by virtue of
\begin{equation}
  \label{eq:momentum2}
  n_{B(F)}(k)=\sum_j \lambda_j |\mu_j(k)|^2\;,
\end{equation}
where $\mu_j(k)$ are the Fourier transforms of the diagonal basis
states. In the special case of free fermions with $\rho_F$ given by
eq.~(\ref{eq:rdspdmf}) these are simply the single particle
eigenstates with $\lambda_j=1$. In Fig.~\ref{fig:momf} we show the
momentum distribution for $N= 19$ and $N=20$ particles for the
potential strengths $\kappa=0$ and $\kappa=20$. One can see that for
the even particle number the introduction of the central
point-potential strongly affects the depth of the oscillations and for
the odd particle number setting it has the effect of smoothing them
out. In this fermionic case the effect can be attributed to the fact
that the point potential introduces non-smooth kinks into the single
particle eigenstates which build the Slater determinant.

\begin{figure}[tbp]
  \includegraphics[width=\linewidth]{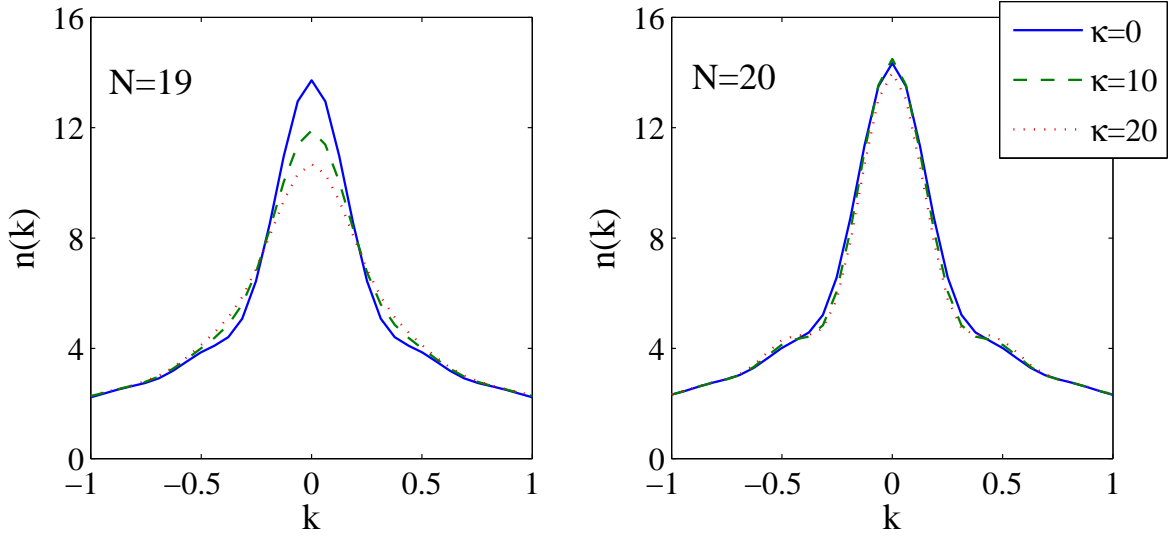}
  \caption{Momentum distribution peak for a $N=19$ and $N=20$ particle
    bosonic TG gas in a harmonic trap with a central attractive point
    potential of strength $ \kappa=0,10$ and $20$.}
\label{fig:closemom}
\end{figure}

The momentum distribution of the bosonic case for $N=19$ and $N=20$
particles is shown in Fig.~\ref{fig:closemom}. One can see that for
the odd number sample the $\delta$-potential has the effect of
lowering the peak of the distribution, indicating a loss of coherence
in agreement with Fig.~\ref{fig:occ}. For the $N=20$ plot one can see
the appearance of bi-modality through the introduction of the
attractive point-potential. This can be interpreted this as single
particle interference arising from the fact that one particle is {\sl
  unpaired} and therefore spatially de-localised over both sides of the
split trap by the point potential.

\subsection{Interference Patterns}
\label{subsect:InterferencePatterns}

\begin{figure}[tbp]
\includegraphics[width=\linewidth]{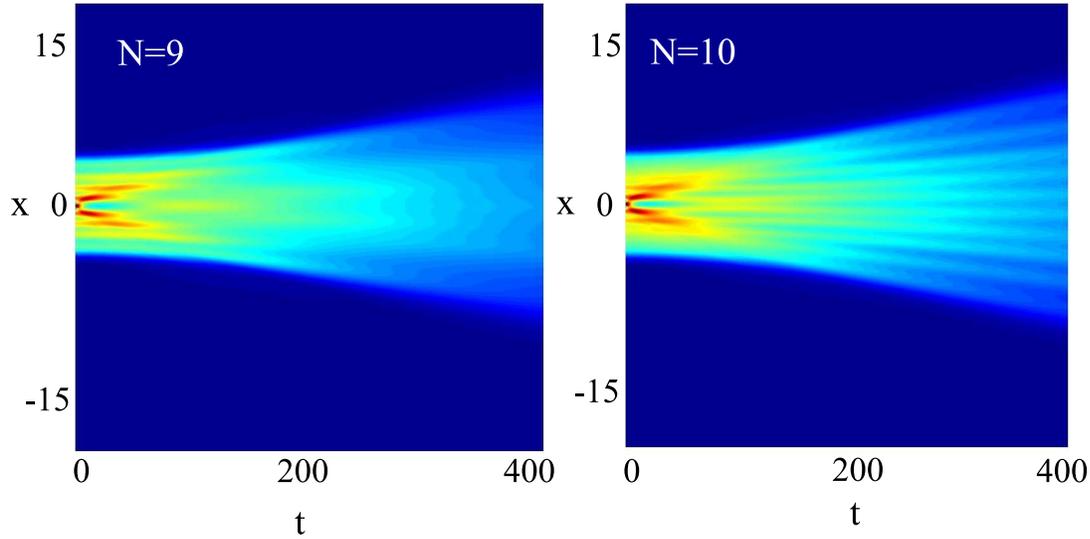}
\caption{Free space time evolution of the single particle density for
  a bosonic TG gas consisting of $N=9$ (left) and $N=10$ (right)
  particles initially in trap with an attractive potential of strength
  $\kappa=30$. Due to the Fermi-Bose mapping the simulations apply
  equally well to fermions.}
\label{fig:interf}
\end{figure}

The Fermi-Bose mapping theorem is also applicable to time dependent
wave-functions dynamics and in this section we will study the time
evolution of the many-body quantum state after removal of all external
potentials. To do this we first find the ground state for the sample
initially confined to the harmonic trap with a central strong
attractive point potential. We then calculate the time evolution of
this state as both, the trap and the central splitting, are turned off
and the gas undergoes free temporal evolution. During this particles
in both halves of the trap will start overlapping and interfering.
The single particle densities for two samples with odd ($N=9$) and
even ($N=10$) particle number are shown in Fig.~\ref{fig:interf}. It
is clearly noticeable that the fringe visibility in the even case is
much higher than in the odd case, where there is virtually none. This
again is a consequence of the larger coherence associated with an even
particle number and is consistent with the results in previous
sections. As we are describing spatial densities here, this effect is
also present in a non-interacting fermion gas. 

\section{Conclusions}
\label{sec:Conclusions}
In this work we undertook a thorough investigation of ground state
properties of 1-D quantum gases in a harmonic trap in the presence of
a point-like attractive potential. While our analysis makes use of an
idealized $\delta$-function potential, it is well known that this
approximation encapsulate the basic, qualitative physics of
experimentally realistic potentials like focused laser beams or
trapped impurities. At these low temperatures, the interaction between
a gas and the later one would be particularly well described by a
point-like potential.

We have studied both, the many-body properties of a bosonic
Tonks-Girardeau and a spin-polarized fermionic gas and calculated the
standard many body quantities such as the single particle densities
and pair distribution functions. The single particle density was found
to be centrally disturbed, in a point-like manner, with effectively no
influence on the overall width if the distribution. The
pair-distribution function also showed an increase in magnitude for
the positions $x,x'=0$ for increasing potential strength and both
these effects could be attributed to the bound eigenstate of the
$\delta$-potential.

We have calculated and shown the reduced density matrices for a range
of particle numbers and potential strength and, by diagonalisation,
were able to derive the the ground state occupation fraction, which is
a measure of the coherence inherent in the gas. It was shown that the
introduction of the potential to the harmonic trap introduced
oscillations in the coherence, with samples consisting of even
particle numbers showing larger values. This was confirmed by
calculating the experimentally accessible quantities of the momentum
distributions and the interference pattern in a time-of-flight
experiment.

With results for both, repulsive as well as attractive point
potentials, available one can envisage an interesting range of
experiments in which the potential strength can be varied as a
function of time. For atomic impurities, for example, this would
simply correspond to driving the inter impurity-gas scattering length
through a Feshbach resonance.

\section{Acknowledgements} 
This project was supported by Science Foundation Ireland under project
number 05/IN/I852. D.O'D.~has been supported through a UREKA grant
05/IN.1/I852ur07.1. J.G.~would like to thank S. McEndoo for
enlightening discussions.

\end{document}